\documentclass[aps,twocolumn,prd,epsf,nofootinbib]{revtex4}

\usepackage{amssymb}
\usepackage{graphicx}
\usepackage{soul}
\usepackage{amsmath}
\usepackage{amsfonts}
\usepackage{comment} 
\usepackage{placeins}
\usepackage{float}
\usepackage{mathrsfs} 

\newcommand{\be}{\begin{equation}}
\newcommand{\ee}{\end{equation}}
\newcommand{\bea}{\begin{eqnarray}}
\newcommand{\eea}{\end{eqnarray}}

\usepackage{color}

\def\beq{\begin{equation}}
\def\eeq{\end{equation}}
\def\baq{\begin{eqnarray}}
\def\eaq{\end{eqnarray}}

\newcommand{\bmp}{\noindent\begin{minipage}{16cm}}
\newcommand{\emp}{\end{minipage}\vskip 7mm} 
\def\lsim{\mathrel{\raise.3ex\hbox{$<$\kern-.75em\lower1ex\hbox{$\sim$}}}}
\def\gsim{\mathrel{\raise.3ex\hbox{$>$\kern-.75em\lower1ex\hbox{$\sim$}}}}

\newcommand{\intron}[1]{}

\def\p{{\partial}}

\begin{document}

\title{ Non-Gaussianity after many-field reheating}

\author{Pedro Carrilho$^1$, David Mulryne$^1$ and John Ronayne$^2$}
\affiliation{$^1$School of Physics and Astronomy, Queen Mary University of London, Mile End Road, London, E1 4NS, UK\\
$^2$Institut d’Astrophysique de Paris, GReCO, UMR 7095 du CNRS et de Sorbonne Universit\'{e}, 98bis boulevard Arago, Paris 75014, France}

\date{\today}
\begin{abstract}
We numerically investigate reheating after quadratic inflation with up to 65 fields, focusing on the production of non-Gaussianity. 
We consider several sets of initial conditions, masses and decay rates. As expected we find that the reheating phase can have a significant effect on the non-Gaussian signal, but that for this number of fields a detectable level of non-Gaussianity requires the initial conditions, mass range and decay rates to be ordered in a particular way. We speculate on whether this might change in the N-flation limit.

\end{abstract}
\maketitle

\section{Introduction}

The theory of inflation has been extremely successful in explaining the initial state of the Hot Big Bang, simultaneously providing a mechanism that generates the initial seeds of structure~\cite{Starobinsky:1980te,Sato:1980yn,Guth:1980zm,Linde:1981mu,Albrecht:1982wi}. Yet the underlying details of the physics of inflation are still far from clear. 
While models of inflation based on the dynamics of a single scalar field slowly rolling on a concave potential are in full 
agreement with all present constraints, the high energies at which inflation occurs 
motivate models with many  light degrees of freedom~\cite{Grana:2005jc,Douglas:2006es}.

In addition to being well-motivated from the fundamental theory point of view, many-field inflation has a rich phenomenology, the consequences of which are not yet fully understood. For example, despite its increased complexity, it has been argued that many-field inflation typically gives simple predictions for observables at the end of inflation that converge statistically towards values in agreement with current experimental constraints~\cite{Frazer:2013zoa,Easther:2013rva,Dias:2016slx,Dias:2017gva,Bjorkmo:2017nzd,Dias:2018pgj} (although dependence on the choice of prior remains \cite{Christodoulidis:2019hhq}). 
In many cases, however, there are still isocurvature 
modes present at the end of inflation~\cite{Easther:2013rva,Dias:2017gva}, which may lead to observables quantities subsequently evolving. Of particular interest is local non-Gaussianity, which remains a key target of upcoming cosmological surveys, \cite{Amendola:2016saw,Camera:2014bwa,Abell:2009aa,Aghamousa:2016zmz,Dore:2014cca}, and an observable level of which requires super-horizon evolution \cite{Maldacena:2002vr, Lyth:2005fi}. 

After inflation, the energy stored in the scalar field(s) is converted to other particles and fields during reheating~\cite{Albrecht:1982mp,Dolgov:1982th,Abbott:1982hn,Kofman:1994rk,Kofman:1995fi,Kofman:1997yn}. In single-field scenarios, this process changes only the time at which the observed scales exit the horizon~\cite{Martin:2014nya,Martin:2016oyk} since the uniform density curvature perturbation, $\zeta$, remains conserved after horizon crossing~\cite{Wands:2000dp,Lyth:2004gb,Christopherson:2014bea,Dias:2014msa,Carrilho:2015cma}.
In contrast, when multiple fields are active during inflation, isocurvature fluctuations may later be converted into curvature fluctuations, thus triggering the super-horizon evolution of $\zeta$. As a consequence, all predictions from multi-field models can be sensitive to the details of reheating when isocurvature modes persist.

Reheating, and the propagation of perturbations through reheating, is in general extremely challenging to understand analytically ~\cite{Greene:1997fu,Meyers:2013gua,Elliston:2014zea,Gonzalez:2018jax} or numerically~\cite{Prokopec:1996rr,Felder:2000hq,Frolov:2008hy,Sainio:2009hm,Easther:2010qz,Huang:2011gf,Sainio:2012mw,Child:2013ria,Nguyen:2019kbm,Lozanov:2019jff,Imrith:2019njf} (see Ref.~\cite{Bassett:2005xm,Amin:2014eta,Lozanov:2019jxc} for reviews). Progress can be made in calculating the observational consequences of many-field reheating, however, by assuming reheating can be modelled as a perturbative process that obeys simple coupled equations. Although a simplification, this approach is self-contained and internally consistent, and  has been taken for small numbers of fields in the past \cite{Elliston:2011et,Elliston:2011dr,Leung:2012ve,Meyers:2013gua}, and recently for many fields by Hotinli {\it et al.}~\cite{Hotinli:2017vhx}. This latter study however considered only two-point statistics that can be mimicked by single field models. In the present paper, we aim to revisit many-field reheating but consider also local non-Gaussianity, which cannot. 

\section{The model}

We consider $N_\phi$ canonical scalar fields, $\phi^I$, with potential $V =\sum_I m^2_{I} \phi^I \phi^I$. 
This model is chosen both to ensure that any observable non-Gaussian signal is purely due to the effects of reheating, and   to make contact with earlier work \cite{Hotinli:2017vhx}. 
Although in principle the different fields could reheat into any number of fluids, we will assume that after reheating 
all species are in thermal equilibrium\footnote{This implies that isocurvature fluctuations~\cite{Bucher:1999re,Carrilho:2018mqy} do not survive after reheating, which is consistent with Planck constraints~\cite{Akrami:2018odb}.}. This 
 is equivalent to having only one effective radiation fluid, $\rho_r$, into which all fields decay.\

In the present study, we make some further simplifying
assumptions that we now discuss. 
Each scalar field falls into the minimum of the potential when the Hubble rate falls below that field's effective mass. If a field reaches the minimum shortly before  inflation ends, or thereafter, it begins to oscillate, and it is only during these oscillations that the 
phenomenological description of perturbative reheating is expected to be accurate. This phase can last for many e-folds during which the frequency of the oscillations grows exponentially. 
Tracking these oscillations numerically can 
therefore be extremely challenging. 
To circumvent this issue, we will treat oscillating fields as matter fluids. This is a good approximation for scalar fields oscillating in a quadratic minimum~\cite{Turner:1983he}. To avoid having to impose a 
transition from oscillating field to matter fluid by hand, we simply introduce as many dust fluids, $\rho_{m(I)}$, as there are fields,
 and introduce decay couplings between each field and its 
 corresponding fluid, effectively making the fields ``decay'' into dust, before these dust fluids decay into radiation. We then remove  
 fields from the system when field's energy density is negligible relative its corresponding dust fluid. These approximations introduce negligible error in predictions while accelerating numerical implementations substantially.
    
With these modifications, our final evolution equations become
\begin{align}
\phi_I''&=-\left(\frac{H'}{H}+3\right)\phi_I'-\frac{\p_I V}{H^2}-\Gamma^{m(I)}\frac{\phi'_{I}}{H}\,,\label{full1}\\
\rho_{m(I)}'&=-3\rho_{m(I)}+H\phi_{I}^{\prime\ 2}\Gamma^{m(I)}-\frac{1}{H}\Gamma^{r}_{m(I)}\rho_{m(I)}\,,\label{full12}\\
\rho_{r}'&=-4\rho_{r}+\frac{1}{H}\sum_I\Gamma^{r}_{m(I)}\rho_{m(I)}\,,\label{full123}
\end{align}
where for later convenience we have written the equations using e-fold number $N$ as the measure of time ($N=\log(a/a_i)$
where $a_i$ is the initial value of the scale factor $a$) and a dash denotes a derivative with respect to $N$. Here 
$\Gamma^{m(I)}$ indicates the decay rate of 
the $I$th field into its corresponding dust fluid, and $\Gamma^{r}_{m(I)}$ the decay rate of this fluid into 
radiation.  $H\equiv \dot{a}/{a}$ is the Hubble rate (where a dot denotes a derivative with respect to cosmic time), and we assume a flat Friedmann-Lema\^{i}tre-Robertson-Walker (FLRW) Universe such that 
\begin{align}
\label{full2}
H^2=&\frac{V(\phi)+\rho_r+\rho_m}{3-\frac12\phi^{\prime\,2}}\,,\nonumber \\
H'=-&\frac12H\phi^{\prime\,2}-\frac{3\rho_m+4\rho_r}{6H}\,,
\end{align}
where we have used $\rho_m$ to represent the sum of the densities of all dust fluids, and $\phi^{\prime\,2}$ the sum of squares of all field velocities. 
 
 Naturally, the effective dust species can only exist once the fields begin oscillating, thus we only activate the decay rates $\Gamma^{m(I)}$ at that time. This makes $\Gamma^{m(I)}$ a function of the fields and fluids present. 
 A sharp transition
 would give rise to pathologies in the perturbation equations, as we will see, 
 and so we parametrise the transition using
\be
\Gamma^{m(I)}=\frac12\bar{\Gamma}^{m(I)}\left(1+\tanh\left[\frac{\pi}{T}\left(1-\frac{H}{H_I}\right)\right]\right)\,
\ee
where $T$ parametrises the width of the transition, $H_I$ is the Hubble rate when the transition happens and $\bar{\Gamma}^{m(I)}$ are the final values of the decay rates after the transition. We choose $H_I=m_I/\sqrt{10}$, where $m_I$ is the mass of field $I$, to ensure that the fields are starting to oscillate and $\bar{\Gamma}^{m(I)}=5m_I$, to guarantee that the fields convert quickly to the dust fluids, which is essential for a fast numerical solution. 
The couplings between dust and radiation, $\Gamma^{r}_{m(I)}$, are also activated in the same way and using the same parametrisation, but with $\bar{\Gamma}^r_{m(I)}$ being chosen according to the case under study. 

\section{Perturbations}

We are interested in studying the evolution of perturbations through reheating, in particular the two and three-point correlation functions of the curvature perturbation, $\zeta$, and the two-point correlations of tensor fluctuations, $h_{ij}$. We  evolve these correlations using the transport approach~\cite{Mulryne:2013uka,Anderson:2012em,Seery:2012vj,Mulryne:2010rp,Mulryne:2009kh,Dias:2011xy,Dias:2014msa,Dias:2015rca,Dias:2016rjq}. This approach takes equations for the evolution of 
perturbations, and uses them to evolve the correlations directly. This can be done using the 
full wavelength-dependent equations from  quantum initial 
conditions on sub-horizon scales, and numerical packages including PyTransport \cite{Mulryne:2016mzv,Ronayne:2017qzn} and CppTransport \cite{Seery:2016lko,Butchers:2018hds} 
implement this approach for models with many fields. Unfortunately neither yet supports reheating, and indeed we are updating  PyTransport with a more general version of a bespoke code developed for this paper. 

To evolve the correlations of interest through reheating, we need evolution equations for perturbations to the fields and 
fluids up to second order in perturbation theory. One approach would be to use full cosmological perturbation theory. However, given that all observable wavelengths are larger than the cosmological horizon at the time of reheating and become wavelength independent, we can instead generate completely equivalent 
perturbation equations using the separate universe approach. This assumes that, on the largest scales, cosmological fluctuations can be modelled by different background patches evolving under the same evolution equations, but having slightly different initial conditions. If different patches are always compared after the same number of e-folds, 
differences in scalar quantities between them (such as values of the field) correspond to 
perturbations on flat hypersurfaces.

\subsection{$\delta N$}
At the end of our evolution the curvature perturbation can be extracted by calculating the perturbative difference in expansion between neighbouring patches needed to reach the same uniform density hypersurface. 
This ``$\delta N$'' quantity 
gives 
the curvature fluctuation on uniform density hypersurfaces, $\zeta$, and can be expanded in terms of the 
field and fluid fluctuations at the end of our evolution as 
\be
\zeta=N_a\delta X^a+\frac12 N_{ab}\delta X^a\delta X^b\,,
\ee
where  $\delta X_a$ represents fluctuation in the fields, field velocities (with respect to e-fold number) and fluids, and 
where a sum over repeated indices is implicit. One form of the tensors $N_a$, $N_{ab}$ is given by
\begin{align}
\label{N1}
N_a = &-\frac{\p_a H}{H'}\,,\\
N_{ab} =& -\frac{\p_a \p_b H^2}{2 H H'}-\p_{(a}(H H')^{-1}\p_{b)}H^2 \nonumber\\
&+\frac{H}{H'}\p_c(H H')^{-1} X^{c\,\prime}\p_a H\p_b H\,,\label{N2}
\end{align}
in which $\p_a$ denotes derivatives with respect to any variable $X_a=\{\phi^I,\,\phi^{I\,\prime},\,\rho_{m(I)},\, \rho_r\}$. These expressions are to be evaluated 
at the end of our evolution using the expressions for the $H$ and $H'$ in Eqs.~\eqref{full2}.
See Ref.~\cite{Dias:2014msa} and references therein for a detailed 
derivation of these coefficients using both the separate universe approach and 
cosmological perturbation theory for the case of multiple fluids, which 
can trivially be extended to include fluids.

\subsection{Evolution on a flat slicing}

Because Eqs.~(\ref{full1}-\ref{full123}) are written using the time variable $N$, together with Eqs.~(\ref{full2}) they 
can be used to derive 
evolution equations for the perturbations 
on flat (constant $N$) hypersurfaces by simply perturbing them up to second order in fluctuations. A 
general variable, $\delta X_a$, evolves according to
\be
\label{evo_perts}
\delta X_a'= u_a^{\ b}\delta X_b+ u_a^{\ bc}\frac12\delta X_b \delta X_c\,.
\ee
with the coefficients given by
\begin{align}
u_a^{\ b}&=\frac{\p X_a'}{\p X_b}\\
u_a^{\ bc}&=\frac{\p^2 X_a'}{\p X_b\p X_c}\,,
\end{align}
where $X_a'$ is expression for the background evolution
of the quantity $X_a$ given by  Eqs.~(\ref{full1}-\ref{full123})
(note the $\phi''$ equation must be split into two first order equations).

In is extremely important to recognise that because the decay rates are time-dependent, 
the evolution equations involve fluctuations of the decay rates
\be
\delta\Gamma\equiv\frac{\p \Gamma}{\p X_b}\delta X_b+\frac12\frac{\p^2 \Gamma}{\p X_b\p X_c}\delta X_b \delta X_c\,,
\ee
that must be present for self-consistency. 
The formalism employed here shows that turning on the $\Gamma$s instantaneously at some time in the perturbation equations (no matter how they are derived) without altering the evolution equations, as has been done in the past, is inconsistent. 

\subsection{Transport}

We define correlations of our perturbations as 
\begin{align}
\langle \delta X_a (k_1)\delta X_b(k_2)\rangle = & (2\pi)^3\delta(k_1+k_2) \Sigma_{ab}(k_1)\,,\\
\langle \delta X_a (k_1)\delta X_b(k_2) \delta X_c(k_3)\rangle = &(2\pi)^3\delta(k_1+k_2+k_3) \nonumber \\
&\times \alpha_{abc}(k_1,k_2,k_3)\,.
\end{align}
where $\Sigma_{ab}$ is the power spectrum of the perturbations and $\alpha_{abc}$ the bispectrum. 
It is straightforward to show, using Eq.~\eqref{evo_perts}, that these spectra evolve according to
\begin{align}
\label{transportevo}
\Sigma_{ab}' &=u_a^{\ c}\Sigma_{bc}+u_b^{\ c}\Sigma_{ac}\,,\\
\alpha_{abc}' &=u_a^{\ d}\alpha_{bcd}+u_a^{\ de}\Sigma_{db}\Sigma_{ec}+\text{perms}\,,
\end{align}
The power spectrum, $P_\zeta$ and bispectrum $B_\zeta$ of the curvature fluctuation are similarly defined as 
\begin{align}
\langle \zeta (k_1)\zeta(k_2)\rangle  &= (2\pi)^3\delta(k_1+k_2) P_\zeta(k_1)\,,\\
\langle \zeta (k_1)\zeta(k_2) \zeta(k_3)\rangle &= (2\pi)^3\delta(k_1+k_2+k_3)B_\zeta(k_1,k_2,k_3)\,,
\end{align}
and are given by
\begin{align}
P_\zeta(k_1)=&\Sigma_{ab}(k_1) N^a N^b\,,\\
B_\zeta(k_1,k_2,k_3)=&\alpha_{abc}(k_1,k_2,k_3) N^a N^b N^c\nonumber \\
				+&\Sigma_{ad}(k_1)\Sigma_{bc}(k_2)N^{ab}N^c N^d+\text{perms}\,,
\end{align}
where $N^a$ and $N^{ab}$ are given in Eqs.~(\ref{N1}-\ref{N2}).

Finally we note that tensor fluctuations are constant on super-horizon scales. 

\subsection{Selecting the scale to evolve}
\label{scale}
If we wished to evolve correlations from quantum sub-horizon scales until the end of reheating, the 
existing implementation of the transport formalism within PyTransport could be employed, stopping the evolution sometime before or shortly after the end of inflation when all observational scales have left the 
horizon. We could then restart the evolution using the output of PyTransport 
as initial conditions for the super-horizon equations (\ref{evo_perts}) that include the reheating 
dynamics. Indeed this is our intention 
for the aforementioned updates to PyTransport. In the present case, however, because the 
inflationary model is 
so simple, it will suffice to use analytic initial conditions for the correlations evaluated 
at horizon crossing (which is the usual procedure in for example the standard $\delta N$ formalism), 
and evolve the 
fluctuations from horizon crossing onwards. Indicating 
quantities at horizon crossing with a $*$, these are given by
\begin{align}
\label{initial}
\Sigma(k^*)_{IJ} &= \frac{H^2_*}{(2\pi)^2}\delta_{IJ}\,,\nonumber\\
\alpha(k^*,k^*,k^*)_{abc} &\approx 0\,,
\end{align}
where all other elements of $\Sigma$ are zero. The tensor spectrum is 
\be
P_T=\frac{4 H_*^2}{k^3}\,.
\ee

Following this procedure the scale, $k$, under consideration can be defined 
purely by the horizon crossing time at which one chooses to set the initial conditions.  
In this project we will be interested in evolving solely the pivot scale used in standard CMB analysis, $k^*=0.05$ Mpc$^{-1}$. Therefore, we require the number of e-folds before the end of inflation at which this scale leaves the horizon ($k^*=a^*H^*$). We use the following formula to relate the pivot scale to the number of e-folds~\cite{Liddle:2003as,Hotinli:2017vhx}:
\be
\label{pivotk}
\frac{k^*}{a_0H_0}=\frac{a^*}{a_{reh}} \frac{a_{reh}}{a_{eq}} \frac{a_{eq}}{a_0}\frac{H^*}{H_0}\,
\ee
where subscript $0$ indicates quantities today, $eq$ quantities at matter-radiation equality and $reh$ quantities at the end of reheating. 
One finds
\be
\frac{a^*}{a_{reh}}=e^{-N^*-\Delta N_{reh}}\,,
\ee
where $N^*$ is the number of e-folds of inflation and $\Delta N_{reh}$ is the duration of reheating. We employ the approximation that between reheating and matter-radiation equality ($a_{eq}$) the universe is radiation dominated, such that $H\propto a^{-2}$ during that stage. This allows us to write
\be
\frac{a_{reh}}{a_{eq}}=\sqrt{\frac{H_{eq}}{H_{reh}}}\,.
\ee
Additionally, the ratio $a_{eq}/a_0$ is simply given by the redshift at equality $a_{eq}/a_0=(1+z_{eq})^{-1}$, and $H_0$ is the measured Hubble constant. The remaining factors in Eq.~\eqref{pivotk} can be tuned to give the correct scale by varying the number of e-folds of inflation and the energy scale of inflation, while also guaranteeing the scalar amplitude, $A_s$, agrees with observations.

\section{Numerical results}

We are now in a position to generate results. 
To solve this system of many fields including reheating, we have developed a python package to evolve Eqs.~(\ref{full1}-\ref{full2}) for the background and Eqs.~\eqref{transportevo} for the perturbations from initial 
conditions in Eq.~\eqref{initial}. We solve the $\Sigma$ equations for 2 length scales close to the pivot scale to 
estimate the spectral index $n_s$ via $n_s=d\log k^3P_\zeta/d\log k$. At this scale 
the spectral index is constrained to be $n_s=0.965\pm0.004$~\cite{Akrami:2018odb}. 
Finally we solve for $\alpha$ for a single scale in order to estimate the local $f_{\text{NL}}$ parameter 
via
\be
f_{\text{NL}}=\frac{5}{18}\frac{B_\zeta(k,k,k)}{P_\zeta(k)^2}\,,
\ee
constrained by Planck to be $f_{\text{NL}}=-0.9\pm5.1$~\cite{Akrami:2019izv}.
We are implicitly assuming that the bispectrum is well approximated by the local shape, as is expected in canonical multi-field models~\cite{Seery:2005gb,Kenton:2015lxa,Kenton:2016abp}. 
In addition we compute the tensor-to-scalar ratio $r$ defined by
\be
r=\frac{P_T}{P_\zeta}\,,
\ee
whose upper bound is $r<0.06$~\cite{Ade:2018gkx}.

\subsection{Curvaton-like set-up}
We consider several different set-ups for our choices of masses, decay rates and initial conditions. 

First we follow Hotinli  {\it et al.}~\cite{Hotinli:2017vhx} and 
consider masses logarithmically sampled in the interval $m_I\in \Lambda [1,10^3]$, where $\Lambda$ is an overall scale to be adjusted to fit the measured amplitude of perturbations, 
choosing the decay rates according to
\be
\label{GaRad}
\bar\Gamma^{r}_{m(I)}=10^{-4} H_{\text{end}} \left(\frac{m_I}{m_{\text{max}}}\right)^\alpha\,,
\ee
where $H_{\text{end}}$ is the value of the Hubble rate at the end of inflation, $m_{\text{max}}$ is the largest mass in the system and $\alpha$ is a parameter controlling the relative hierarchy of the different decay rates. For the background fields the initial conditions are set in the same order as the masses in the interval $\phi_I\in[10^{-3},20]$, while field derivatives are set using the slow-roll approximation $\phi_I'=-\p_I V /V$. 

These initial conditions select the trajectory in field space followed by the fields, but the initialisation 
of field perturbations are only set a number of e-folds before the end of inflation, $N^*$. This number is determined by approximating the scale under evaluation, $k^*$ to be the pivot scale using the method described in Section~\ref{scale} via a search algorithm, with the overall scale of the potential, $\Lambda$, selected to ensure 
$P_\zeta$ is in agreement with observation.  

The choice of aligning the hierarchy of masses with that of decay rates allows the lightest fields to oscillate for longer, thus resulting in a lengthy dust-dominated stage during reheating, and the choice of initial 
conditions ensures the long lived fields are close to the minimum initially. This enhances curvaton(s)-like behaviour~\cite{Lyth:2001nq,Enqvist:2001zp,Lyth:2002my,Bartolo:2003jx,Malik:2006pm,Assadullahi:2007uw,Alabidi:2010ba,Suyama:2011pu,Sueiro:2012fq,Fonseca:2012cj,Vennin:2015vfa}, such as the production of isocurvatures and later conversion into curvature perturbations, as well as the consequent generation of relatively large non-Gaussianity. Hence we expect this set up to give the most dramatic effects, and it can be seen as a many-field curvaton model.  

\begin{figure}
	\centering
	        \includegraphics[width=0.49\textwidth]{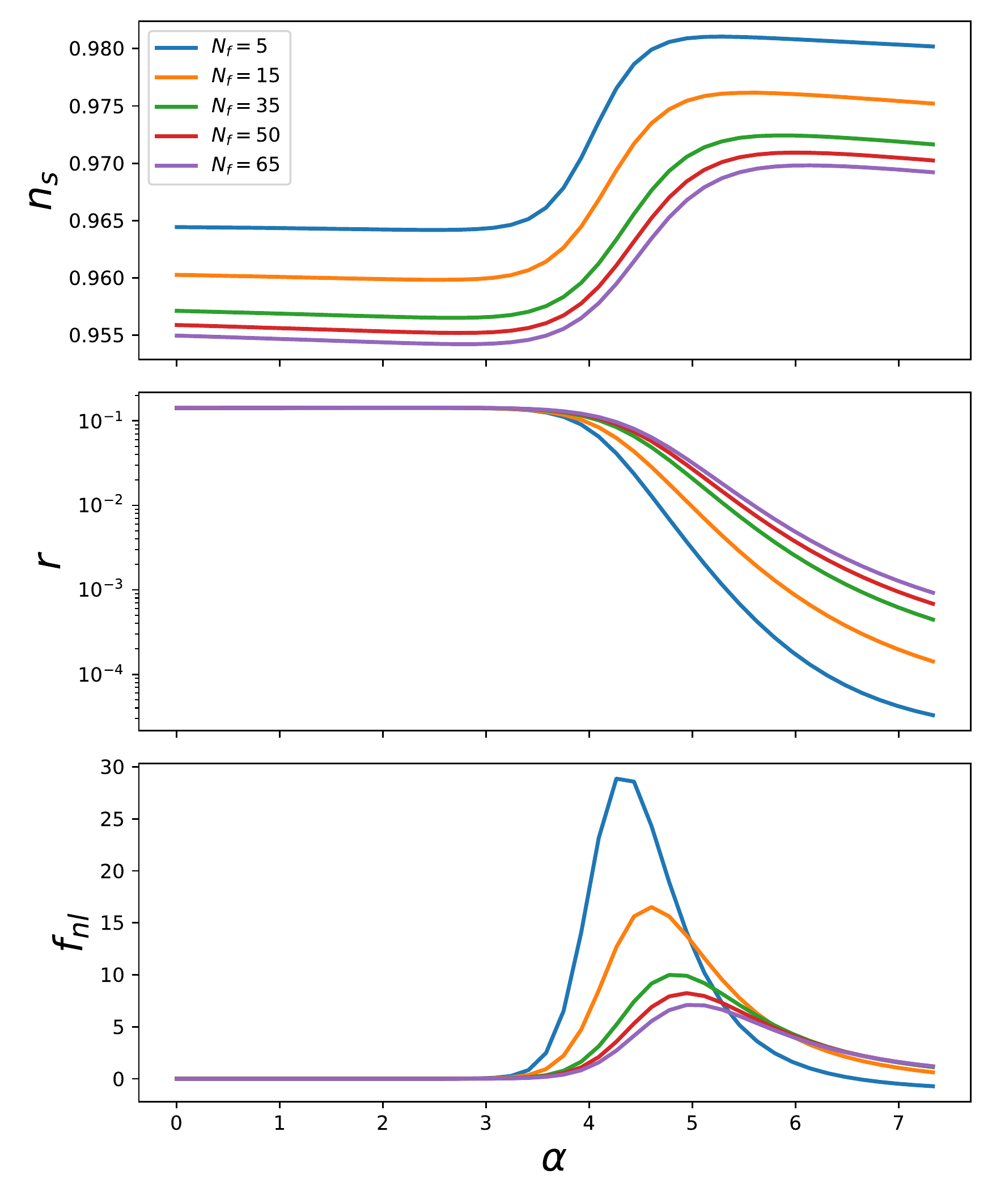}
	\caption{Numerical results for $n_s$ (top), $r$ (middle) and $f_{\text{NL}}$ (bottom), as a function of the parameter $\alpha$ in the original set-up, using Eq.~\eqref{GaRad}. Different lines correspond to different values of the number of fields, as labelled in the legend.}\label{fig:many}
\end{figure}

The results given in Fig.~\ref{fig:many} show the dependence of $n_s$, $r$ and $f_{\text{NL}}$ on the parameter $\alpha$ and the number of active fields $N_f$. We qualitatively reproduce the results of Ref.~\cite{Hotinli:2017vhx} in $n_s$ and $r$. In addition to this, we find that in this set-up, the non-Gaussianity parameter, $f_{\text{NL}}$, peaks at certain values of $\alpha$, decaying into negative values for large $\alpha$. This is similar to what occurs in the curvaton scenarios with all fields becoming massive before they decay (as in cases 4, 5 and 6 described by Ref.~\cite{Vennin:2015vfa}). The parameter $\alpha$ determines the constant temporal separation (in e-folds) between each decay. Thus, for very small $\alpha$ all decays happen close together, such that most of the contribution from the curvature perturbation is from the inflaton(s). This results in observables that are close to those obtained when reheating is neglected. As $\alpha$ increases to $\sim 4$, the decays happen sufficiently apart for some of the isocurvature perturbations to contribute substantially to the curvature perturbation, while their corresponding energy densities are small when they decay, generating large non-Gaussianity and increasing $n_s$ towards scale invariance. Further increase of $\alpha$ induces the initially sub-dominant fields to dominate the total energy density upon their decay, which reduces non-Gaussianity until it reaches the asymptotic value of $f_{\text{NL}}=-5/4$. As the number of fields increases, the different effects arise from multiple fields, instead of having one ``inflaton'' and one ``curvaton''. With many fields, an approximate continuum between inflaton-like fields and curvaton-like fields exists, with those in an intermediate category not contributing substantially to modify the single-field-like behaviour. This makes it more difficult to generate large non-Gaussianity. In addition, since the mass distribution is the same for all numbers of fields, the time separation between decays shortens as $N_f$ increases, thus requiring larger $\alpha$ to get similar effects, explaining the deviation of the peak in $f_{\text{NL}}$. 

Regarding comparison with observations, our results show that low values of $\alpha$ are not compatible with the constraint on $r$, while low numbers of fields are in conflict with the $n_s$ constraint, given the required large value of $\alpha$. Our non-Gaussianity results again imply that only sufficiently large values of $\alpha$ or large $N_f$ are in agreement with current constraints on $f_{\text{NL}}$.

\subsection{Modified set-ups}

We now study modifications from the first set-up chosen above with initial conditions given by $\phi_I\in[10^{-3},20]$, masses by $m_I/\Lambda\in[1,10^3]$ and decay constants by Eq.~\eqref{GaRad}. 

We began by modifying the pre-factor in Eq.~\eqref{GaRad} ranging from $10^{-2}$ to $10^{-8}$, finding negligible variation of the results with respect to those obtained with $10^{-4}$. This is well explained by the interpretation outlined above and shows again that the time separation between decay times, parametrized by $\alpha$, is the most relevant parameter to explain the variation of observables.

We also explored scenarios in which the dependence of the decay constants, $\Gamma$, on mass is motivated by the type of particle into which the scalar fields decay. It well known that decay into scalar particles results in $\Gamma\propto m^{-1}$, while decay into fermions gives $\Gamma\propto m$ \cite{Kofman:1997yn}. While the latter case is already contained in Eq.~\eqref{GaRad} by choosing $\alpha=1$, the case $\Gamma\propto m^{-1}$ is not, so we explore it separately by modifying that equation into
\be
\label{GaRadNeg}
\bar\Gamma^{r}_{m(I)}=10^{-4}\beta H_{\text{end}} \left(\frac{m_I}{m_{\text{min}}}\right)^{-1}\,,
\ee
where $\beta$ is an additional parameter varied between $10^{-2}$ and $10^2$ and the minimum mass is now used instead of the maximum mass to guarantee that $\Gamma^r_{m(I)}<H_{\text{end}}$. This inversion implies that the order of decays is also inverted, such that the lightest fields now decay first and essentially do not contribute to the curvature perturbation. This case then mimics the results obtained with low values of $\alpha$ in the first set-up shown above, with no specific multi-field effects appearing. Similarly to the previous set-up, varying the overall factor $\beta$ does not change results substantially. Varying the exponent from the physically motivated value of $-1$ to more negative values also does not change observables substantially, since broadening the distribution of decay times simply dilutes multi-field effects further. The inversion of decay times was also effectively investigated in Ref.~\cite{Hotinli:2017vhx}, in which the authors randomized the order of decay times and found that nearly only the original ordering generated multi-field effects.

\begin{figure}
	\centering
					\includegraphics[width=0.49\textwidth]{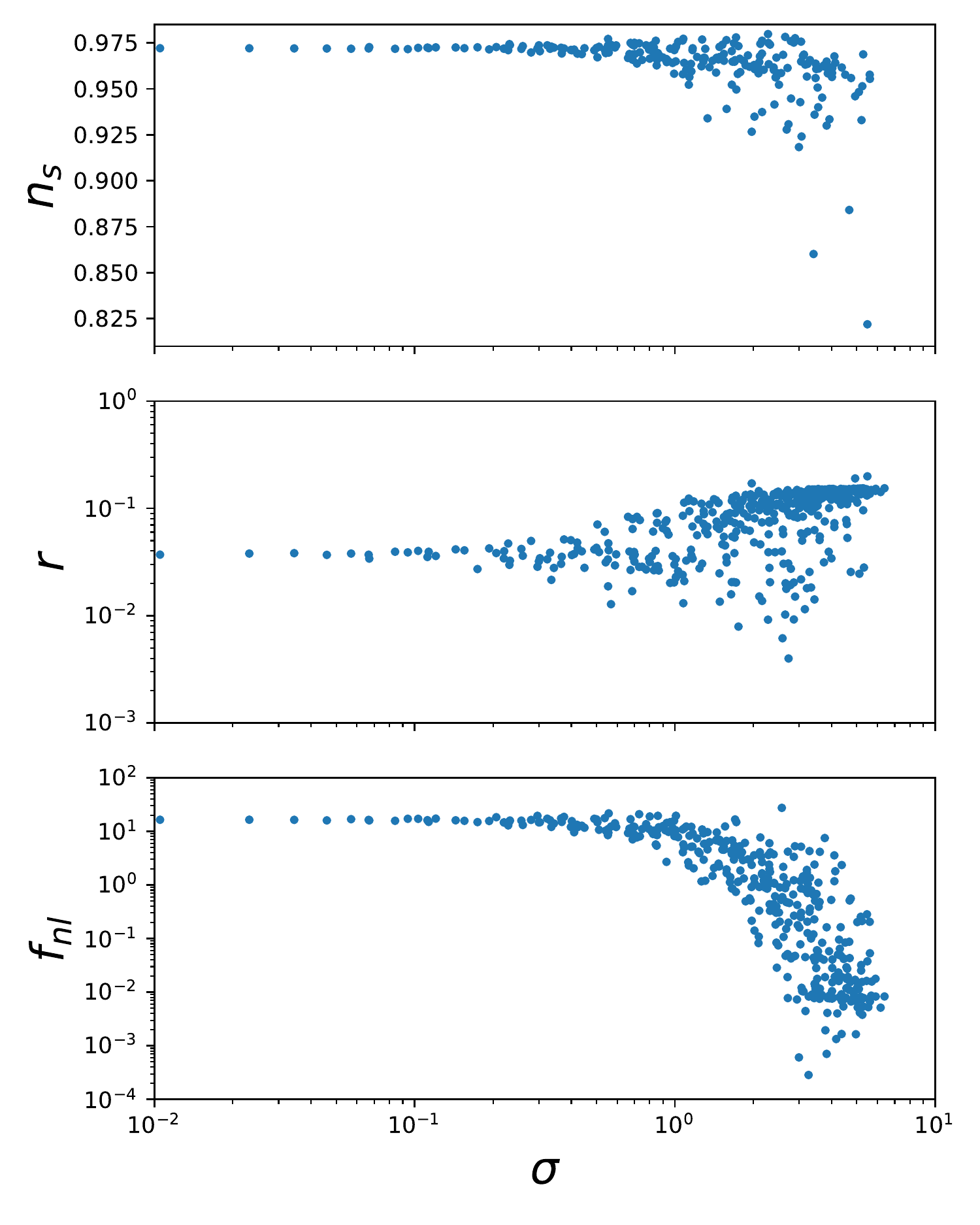}
	\caption{Numerical results for $n_s$ (top), $r$ (middle) and $f_{\text{NL}}$ (bottom), as a function of the dispersion, $\sigma$, around the original initial conditions, for 15 fields and $\alpha=4.5$.}\label{fig:sigma}
\end{figure}

The final modifications we study involve modifying initial conditions away from $\phi_I\in[10^{-3},20]$. We begin by varying the initial condition for each field independently according to a Gaussian distribution in $\log\phi_I$ with standard deviation $\sigma_I$, which we vary in the range $[10^{-2},7]$. We plot the results in Fig.~\ref{fig:sigma}.

We find that deviations away from the original set-up progressively eliminate multi-field effects, reaching, on average, similar results to those of an equivalent single-field model. While the results shown in Fig.~\ref{fig:sigma} are for 15 fields, the trend presented there is also observed for different numbers of fields, with the variance in observables decreasing as a function of number of fields. For small $N_f$, it is thus possible that one finds initial conditions that also give enhanced multi-field effects, such as larger $f_{\text{NL}}$ than in the original scenario. However, these cases are rare and become rarer with number of fields, thus demonstrating that multi-field effects are generally generated only in fine-tuned situations, at least for the model under study. This confirms some of the conclusions of Ref.~\cite{Hotinli:2017vhx} regarding the necessity of fine-tuning, extending them to non-Gaussianity.

\begin{figure}
	\centering
					\includegraphics[width=0.49\textwidth]{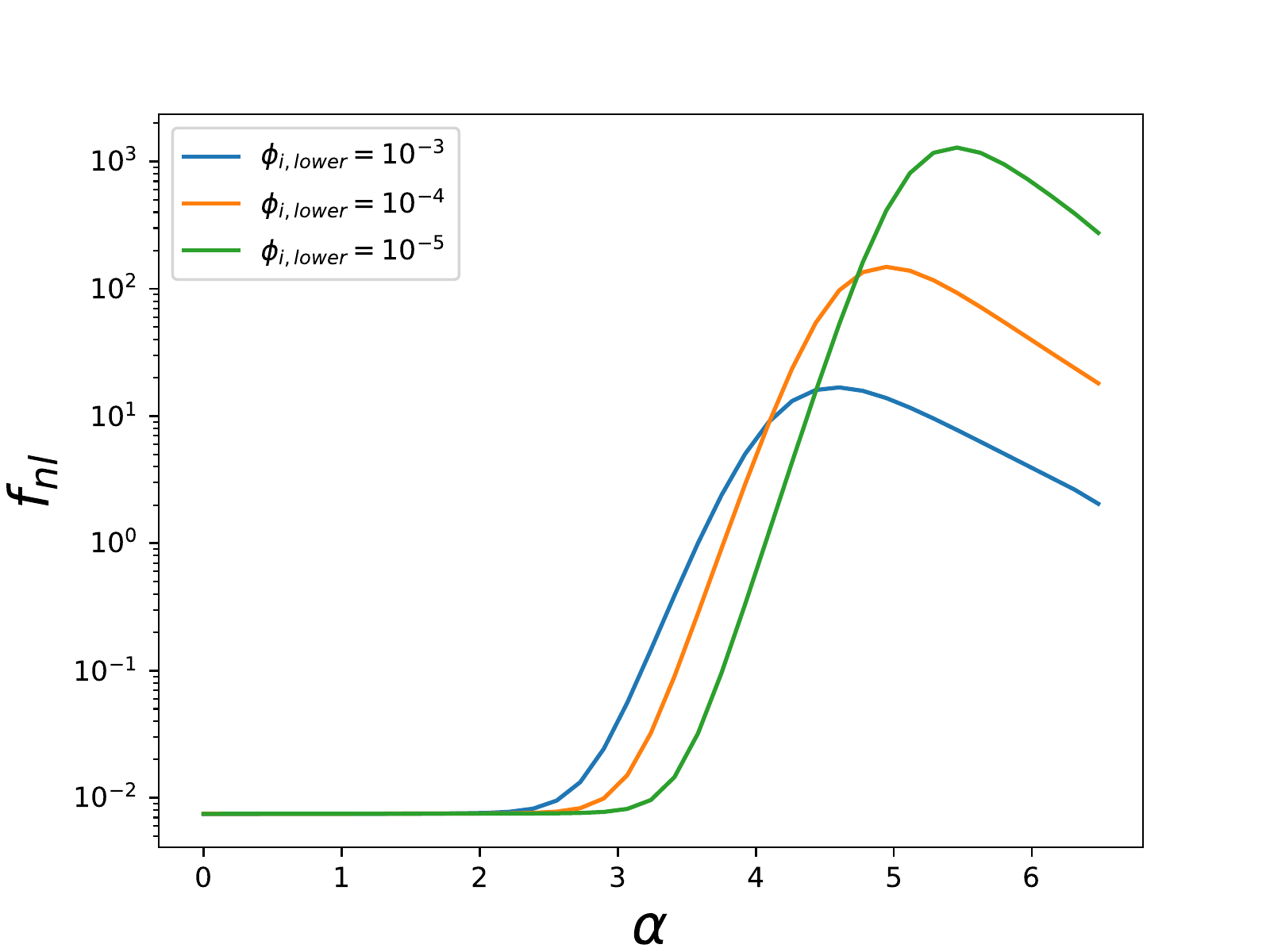}
	\caption{Numerical results for $f_{\text{NL}}$, as a function of $\alpha$, for 15 fields, varying the lower value of the range of initial conditions, $\phi_{\rm{lower}}$.}\label{fig:philower}
\end{figure}

Next, we explore the effects of increasing the range of initial conditions to include smaller values, i.e. we set $\phi_I\in[\phi_{\rm{lower}},20]$. In Fig.~\ref{fig:philower}, we show the results for local non-Gaussianity for different values of the lower limit of the initial condition range, $\phi_{\rm{lower}}$. We see that the peak value of $f_{\text{NL}}$ shifts towards larger $\alpha$ and increases proportionally to $\phi_{\rm{lower}}^{-1}$, to good approximation. These variations are explained by the fact that decreasing the initial conditions of the lightest fields reduced their contribution to the energy density initially. This implies that a larger value of $\alpha$ is required to ensure that lighter fields give a substantial contribution to $\zeta$. However, at the point in which this is attained, the effects on non-Gaussianity are larger, exactly because the contribution of these field to the energy density is smaller. This is again in agreement with the view that this model represents a many-field curvaton, as this is precisely what occurs in the curvaton case.

\section{Discussion and conclusions}

In this paper we have investigated the effect of perturbative reheating after quadratic inflation with many fields. 
Our focus has been on the non-Gaussian signal 
generated by many-field reheating, which complements earlier work at the level of the power spectrum by Hotinli {\it et al.}~\cite{Hotinli:2017vhx}. In our work, we considered up to 65 fields and we found that an observable non-Gaussian signal is consistently 
found only in the case where long-lived fields, which decay at the end of the evolution, are initially close to their minimum (by this we mean a 
field displacement much less that the Planck mass). We found that the energy density in these late decaying fields needed to grow significantly with respect to the radiation produced by other decays for them 
to contribute to $\zeta$ and produce a non-Gaussian signal. In this sense the set-up is required to be like 
 a many-field curvaton model, and we found that as we moved away from this set-up the non-single-field-like behaviour disappeared in the majority of runs. We also found that keeping the range of initial conditions the same, 
 but introducing more fields and hence more densely populating the space of initial conditions decreased the amount of non-Gaussianity produced. While extending the initial condition range to include initial conditions closer to the fields' minima increases
the amount of non-Gaussianity produced.

A motivation for our work is the question of how common non-Gaussianity is in multi-field inflation. This is  important 
to address in the light of the 
hunt for local non-Gaussianity in future surveys, and how any signal that is observed, or not observed, should be interpreted. 
It is well known that producing non-Gaussianity during inflation in models with a small number of 
fields requires features such as hill-tops and ridges to be present, and also requires particular initial conditions 
close to such features that could be considered finely-tuned~\cite{Byrnes:2008wi,Elliston:2011et,Elliston:2012wm}. 
Moreover, in many-field models where the potential is randomly generated a non-Gaussian signal appears to be rare~\cite{Frazer:2011tg,Frazer:2011br,Bjorkmo:2017nzd}.
On the other hand, in models like axion inflation, although the initial conditions needed to produce non-Gaussianity look 
finely tuned when only a few fields are considered, in the N-flation limit where very many fields 
all contribute to sustain inflation such initial conditions become typical~\cite{Kim:2010ud,Kim:2011jea}.
Turning to reheating, in quadratic potentials it would be interesting to ask whether something 
similar happens when the N-flation limit is reached; i.e. when inflation can be sustained 
with all fields initially close to their minima, how common is non-Gaussianity if a range of reheating times is considered? 
To answer this would require simulations with more than a thousand fields, which is computationally 
prohibitively expensive with our current approach and so beyond the scope of our paper. In our study, however, we 
have already seen that non-Gaussianity decreases in this model when the initial condition space is densely populated, and 
that fields closest to the minimum also need to be the ones that decay latest. This is suggestive that non-Gaussianity is not generic 
even in the N-flation limit, but further study of this possibility is needed to provide a definitive answer.

\section*{Acknowledgements}

PC is supported by the STFC under grant ST/P000592/1. DJM is supported by a Royal Society University Research Fellowship.  JWR is supported by the European Research Council under the European Union’s Horizon 2020 research and innovation programme (grant agreement No 758792, project GEODESI). PC and DJM acknowledge the hospitality of the IAP in which a substantial part of this work was initiated, as well as the organization of the Inflation and Geometry workshop. We thank S\'ebastien Renaux-Petel for useful discussions.

\bibliography{Reheat_bib}{}

\end{document}